\documentstyle[11pt,newpasp,twoside,epsf]{article}
\def\edcomment#1{\iffalse\marginpar{\raggedright\sl#1\/}\else\relax\fi}
\reversemarginpar

\newcommand{\hii}{HII}
\newcommand{\uchii}{UC~HII}
\newcommand\hrrl{{H76$\alpha$}}

\begin{document}
\title{Hierarchical Structure of HII Regions - \\ 
Ultracompact, Compact, and Extended Components}
\author{Kee-Tae Kim and Bon-Chul Koo}
\affil{Astronomy Program, SEES, Seoul National University, Korea}
%\author{Ima Co-Author}
%\affil{The Name of My Institution, The Full Address of My Institution}

\begin{abstract}
We have observed extended emission physically associated with 
15 out of 16 \uchii\ regions with large ratios of single-dish to VLA fluxes. 
We discuss their implications for the age problem of 
\uchii\ regions and the evolution of HII regions.
\end{abstract}

%\section{Introduction}

Recently extended emission was observed around a few tens \uchii\ regions
(Koo et al. 1996; Kurtz et al. 1999), suggesting that
the ultracompactness of \uchii\ regions may arise from the missing
short baseline of interferometric observations.
However, the relationship between \uchii\ regions
and the surrounding extended emission is poorly understood at present.

%\section{Observations}
%\medskip
 
In order to address this issue, we carried out radio continuum and 
radio recombination line (RRL) observations of 16 \uchii\ regions,
which were selected from the catalog of Wood \& Churchwell (1989)
because of their simple morphology and large ($>$10) ratios of
single-dish to VLA fluxes.
The radio continuum observations were made with $\sim$30$''$ resolution 
at 21~cm using the VLA DnC-array.
The \hrrl\ RRL observations were undertaken
using the Green Bank 43~m telescope (FWHM$\simeq$2$'$).
We also mapped molecular clouds associated with the 16 \uchii\ regions in
the $^{13}$CO J=1$-$0 and CS J=2$-$1 lines
with the NRAO 12~m (FWHM=60$''$) and the TRAO 14~m (FWHM=60$''$) telescopes,
respectively.
%$^{13}$CO J=1$-$0 and CS J=2$-$1 line observations were made with 
%the NRAO 12 m (FWHM=60$''$) and the TRAO 14 m (FWHM=60$''$) telescopes, 
%respectively.

%\section{Results and Discussion}
%\medskip

We detected extended emission towards all our sources.
The extended emission consists of one to several compact
($\sim$1$'$ or 0.5$-$5~pc) components and a diffuse extended 
envelope (Fig. 1$a$).
All the \uchii\ regions but two are located at the peaks of 
the compact components.
We derived the ratios of single-dish to VLA fluxes
for 52 \uchii\ regions with simple morphology in the catalogs of
Wood \& Churchwell (1989) and Kurtz et al. (1994),
and found that most of them have large ratios.
Therefore, most \uchii\ regions are likely to
be associated with extended emission as our sources
(see Kim \& Koo 2001a for details).

We have found no significant velocity difference
among the \uchii\ region, compact component(s), and extended envelope 
in all the sources except one.
This indicates that the three components are physically associated.
The \uchii\ regions correspond to the peaks of
their associated compact components, as noted above, and
the compact components with \uchii\ regions
are more compact than those without \uchii\ regions.
If the two are ionized by separate sources, one would not expect
these correlations.
Hence,
the \uchii\ region and its associated compact component in each
source are likely to be excited by the same ionizing source.

\vskip -0.05cm
Recent high-resolution molecular line studies of massive star-forming 
regions have revealed `molecular clumps' and `hot cores' therein
(e.g., Cesaroni et al. 1994).
The hot cores are believed to be the sites of massive star formation.
The sizes of the hot cores and molecular clumps agree roughly with
those of \uchii\ regions and their associated compact components, respectively.
Based on these observations,
we propose a simple model in which
the existence of the extended
emission around \uchii\ regions
can be explained by combining the Champagne flow model with
the hierarchical structure of massive star-forming regions 
(Fig. 1$b$).
Our molecular line observations show that $^{13}$CO cores are~associated 
with the compact components regardless of the presence of
\uchii\ regions, while CS cores are preferentially associated with 
the compact components with \uchii\ regions. 
This strongly suggests that the compact components with \uchii\ regions 
are in an earlier evolutionary phase than those without \uchii\ regions. 
By comparing molecular line data with radio continuum and RRL data,
we found champagne flows in 10 sources in our sample (Fig. 1$a$).
These observations are consistent with our model
(see Kim \& Koo 2002 for details).

\begin{figure}
\plottwo{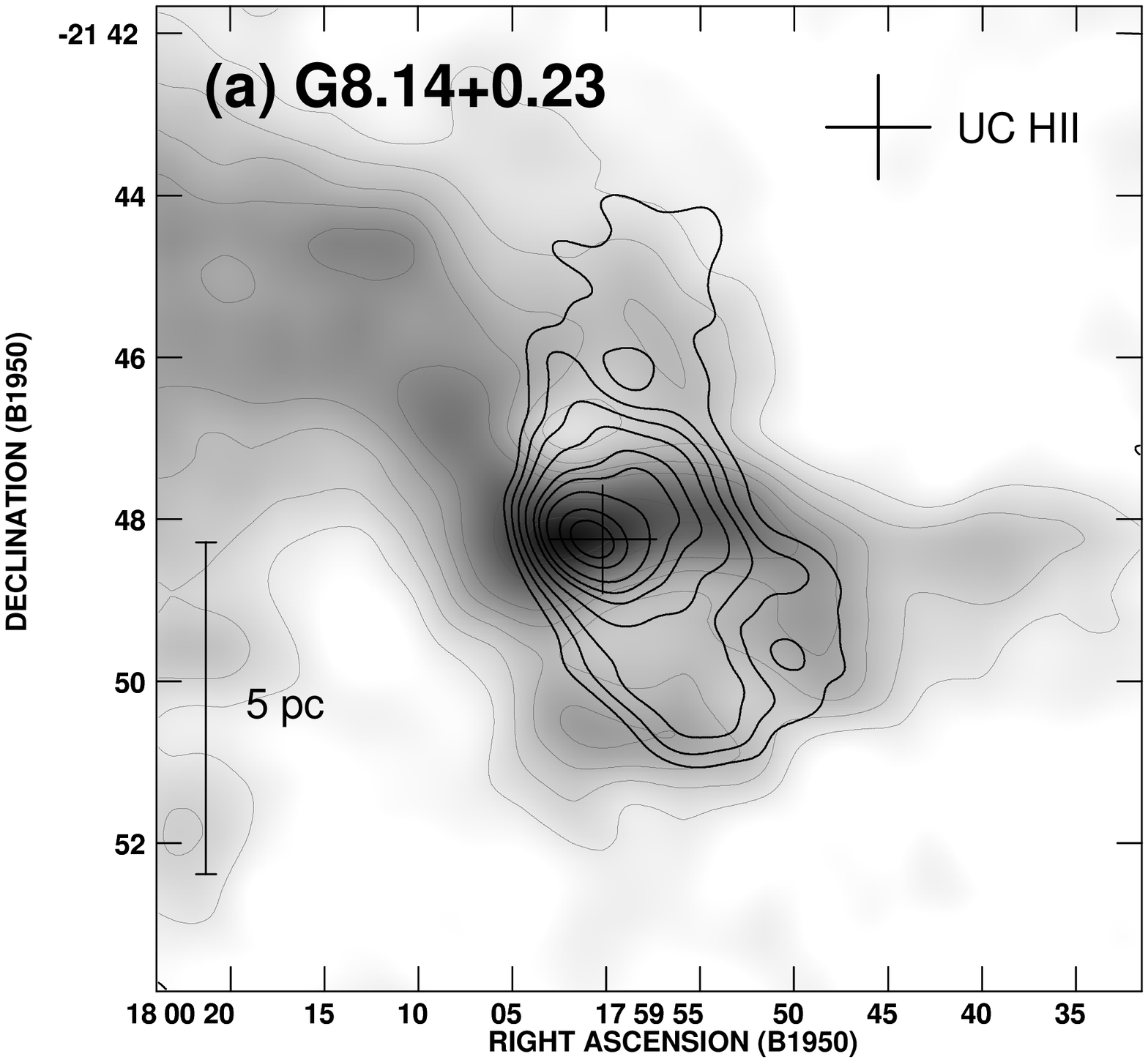}{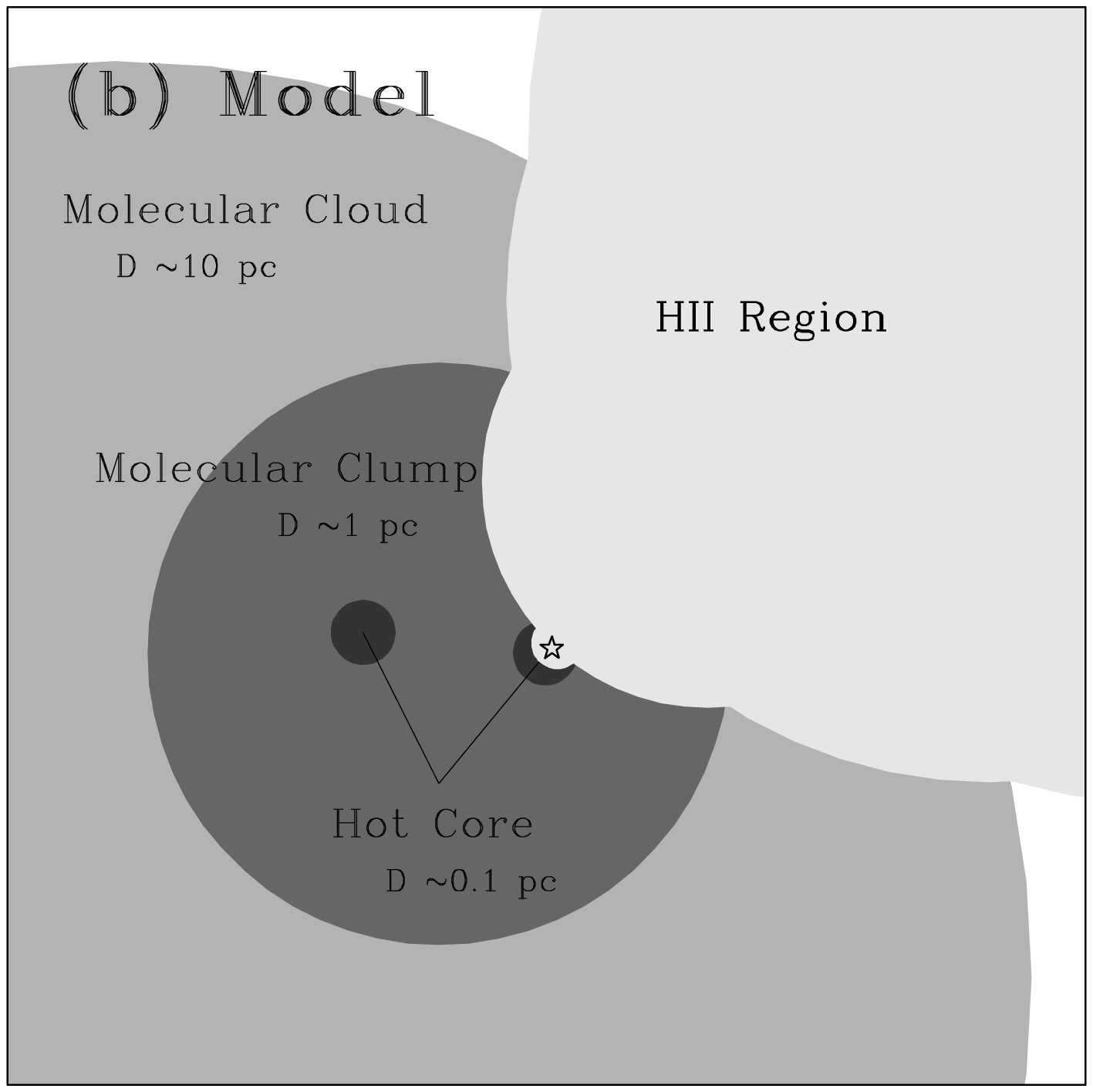}
\vskip -0.6cm
\caption {(a) Radio continuum (contours) and $^{13}$CO line images
for G8.14+0.23; (b) Schematic model (see text)}
\end{figure}

%\medskip

In conclusion,
most sources known as \uchii\ regions may not be ``real'' \uchii\ regions 
but ultracompact cores of more extended \hii\ regions.
Therefore, the ``age problem'' of \uchii\ regions does not seem to be as
serious as earlier studies argued.
The ultracompact, compact, and extended components of HII
regions may not represent an evolutionary sequence
and could coexist for a long ($>$10$^5$ yr) time.
Such a structure of HII regions is likely to be closely related to the
hierarchical structure of parental molecular clouds.

%{\bf Acknowledgements.} 
%\acknowledgements 
%This work was partly supported by the BK21 program,
%Ministry of Education, Korea, through SEES.

\end{document}